\documentclass[twocolumn,showpacs,amsmath,amssymb,nofootinbib,superscriptaddress]{revtex4}
\usepackage{amssymb}
\usepackage{bbm}
\usepackage{mathrsfs}
\usepackage{amsfonts}
\usepackage{mathptmx}
\usepackage{epsfig}
\usepackage{graphicx}
\usepackage{slashed}
\usepackage{color}

\def\comment#1{}

\begin{document}

\title[]{Nonlinear Breit-Wheeler process in the collision of a photon with two plane waves}

\author{Yuan-Bin Wu}
\email{wuyb@icranet.org}%
\affiliation{Department of physics and ICRA, Sapienza University of
Rome, Piazzale Aldo Moro 5, I-00185 Rome, Italy}%
\affiliation{ICRANet, Piazza della Repubblica 10, I-65122 Pescara,
Italy}%
\affiliation{ICRANet, University of Nice-Sophia Antipolis, 28 Avenue
de Valrose, 06103 Nice Cedex 2, France}
\author{She-Sheng Xue}
\email{xue@icranet.it} %
\affiliation{Department of physics and ICRA,
Sapienza University of
Rome, Piazzale Aldo Moro 5, I-00185 Rome, Italy}%
\affiliation{ICRANet, Piazza della Repubblica 10, I-65122 Pescara,
Italy}%


\begin{abstract}

The nonlinear Breit-Wheeler process of electron-positron pair
production off a probe photon colliding with a low-frequency and a
high-frequency electromagnetic wave that propagate in the same
direction is analyzed. We calculate the pair-production probability
and the spectra of the created pair in the nonlinear Breit-Wheeler
processes of pair production off a probe photon colliding with two
plane waves or one of these two plane waves. The differences of
these two cases are discussed. We evidently show, in the two-wave
case, the possibility of Breit-Wheeler pair production with
simultaneous photon emission into the low-frequency wave and the
high multiphoton phenomena: (i) Breit-Wheeler pair production by
absorption of the probe photon and a large number of photons from
the low-frequency wave, in addition to the absorption of one photon
from the high-frequency wave; (ii) Breit-Wheeler pair production by
absorption of the probe photon and one photon from the
high-frequency wave with simultaneous emission of a large number of
photons into the low-frequency wave. The phenomenon of photon
emission into the wave cannot happen in the one-wave case. Compared
with the one-wave case, the contributions from high multiphoton
processes are largely enhanced in the two-wave case. The results
presented in this article show a possible way to access the
observations of the phenomenon of photon emission into the wave and
high multiphoton phenomenon in Breit-Wheeler pair production even
with the laser-beam intensity of order $10^{18}~\rm{W/cm^2}$.

\end{abstract}



\maketitle

\section{Introduction}
\label{sec:intro}

Owing to recent advanced laser technologies, the intensity of laser
beams has been and will be increased several orders of magnitudes in
the last decade  and in the future; it becomes accessible to observe
the fundamental phenomena of the quantum electrodynamics (QED) in
the nonlinear regime of strong fields \cite{Mourou2006, Piazza2012}.
The characteristic scale of strong fields is given by the critical
field $E_c = m^2 c^3/|e|\hbar=1.3\times10^{16}\text{V/cm}$,
corresponding to the critical intensity $I_c \simeq 4.6 \times
10^{29}~\rm{W/cm^2}$ of laser beams. At the critical field, one of
these fundamental phenomena is electron-positron ($e^{-}e^{+}$) pair
production from the QED vacuum that was studied by Sauter, Euler,
and Heisenberg in terms of the Euler-Heisenberg effective Lagrangian
\cite{Sauter1931, Heisenberg1936} and by Schwinger in the QED
framework \cite{Schwinger1951} (for more details see reviews
\cite{Dunne2006, Ruffini2010}).

There are many experiments used to reach the critical field in
ground laboratories: x-ray free electron laser (XFEL) facilities
\cite{XFEL}, optical high-intensity laser facilities such as the
Petawatt laser beam \cite{petawattlaser} or ELI \cite{ELI}, and the
SLAC-E-144 experiment using nonlinear Compton scattering
\cite{Burke1997, Bamber1999} (for details see review articles
\cite{Mourou2006, Salamin2006}). This leads to the physics of
ultra-high-intensity laser-matter interactions near the critical
field (see, e.g., Refs.~\cite{Piazza2012, Dunne2009r, Gies2009,
Kleinert2013, Kohlfurst2014}).

Another nonlinear QED phenomenon whose relevant observations might
be more easily accessible in ground laboratories is the nonlinear
Breit-Wheeler process
\begin{equation}\label{one-wave}
  \gamma'+ \bar{n} \gamma \rightarrow e^{-} + e^{+}.
\end{equation}
The creation of an $e^{-}e^{+}$ pair in the collision of two real
photons $\gamma' + \gamma \rightarrow e^{-} + e^{+}$ was first
considered by Breit and Wheeler \cite{Breit1934}. This linear
Breit-Wheeler process refers to a perturbative QED process. In the
seminal works by Reiss \cite{Reiss1962} and others
\cite{Nikishov1964, Nikishov1965, Nikishov1967, Narozhnyi1965,
Ritus1985}, the authors fully analyzed and discussed the
generalization of the Breit-Wheeler process to the nonlinear process
(\ref{one-wave}) of $e^{-}e^{+}$ pair production off a probe photon
colliding with an intensive monochromatic plane wave. They have
shown that the nonlinear Breit-Wheeler process is a multiphoton
process.

In order to create an $e^{-}e^{+}$ pair, the center-of-mass (CM)
energy of the scattering photons must be larger than the kinematic
energy threshold $2m c^2 \simeq 1.02~\rm{MeV}$, where $m$ is the
electron mass. The phenomenon of pair production in multiphoton
light-by-light scattering of the nonlinear Breit-Wheeler process has
been detected in the SLAC-E-144 experiment \cite{Burke1997,
Bamber1999}. In this experiment, a laser beam with an intensity of
$I\sim 10^{18}~\rm{W/cm^2}$ and a $527~\rm{nm}$ wavelength was
applied. The high-energy probe photon was created by the
backscattering of the laser beam off a high-energy electron beam of
energy $46.6~\rm{GeV}$. With a laser beam of photon energy
$2.35~\rm{eV}$ and probe photons with a maximum energy
$29.2~\rm{GeV}$, a highly nonlinear phenomenon of the Breit-Wheeler
multiphoton process was observed \cite{Burke1997, Bamber1999}.
$e^{-}e^{+}$ pair production by real photons of the Breit-Wheeler
process is one of most relevant elementary processes in high-energy
astrophysics, and it can lead to observable effects such as the
high-energy $\gamma$ spectra and relevant information about the
cosmic background radiation (see, e.g., Refs.~\cite{Ruffini2010,
Hauser2001, Aharonian2003}).

The generalization of this nonlinear Breit-Wheeler multiphoton
process in a monochromatic plane wave of infinitely long pulse to
the case of a finite pulse has been investigated in
Refs.~\cite{Heinzl2010, Titov2012, Nousch2012, Krajewska2012,
Titov2013}, considering the fact that the upcoming intensive optical
laser beams are expected to be very short with only a few
oscillations of the electromagnetic field in their pulses. It has
been shown that the pulse shape and the pulse duration have a
variety of effects on the nonlinear Breit-Wheeler process
\cite{Heinzl2010, Titov2012, Nousch2012, Krajewska2012, Titov2013},
such as the enhancement of the pair-production rate in the
subthreshold region \cite{Titov2012} and the carrier-envelope phase
effects on the distribution of created pairs \cite{Krajewska2012}.

The approach of using two electromagnetic waves and multiple
colliding electromagnetic waves is the future direction of studying
the fundamental QED phenomena in the nonlinear regime of strong
fields, due to its various advantages such as the enhancement of the
pair-production rate (see, e.g., Refs.~\cite{Piazza2012,
Schutzhold2008, Bulanov2010}). As proposed in
Refs.~\cite{Schutzhold2008, Dunne2009}, the Schwinger mechanism of
vacuum pair production is catalyzed by superimposing a
high-frequency beam with a strong low-frequency laser pulse. With a
similar idea, tunneling $e^{-}e^{+}$ pair creation was shown to be
observable with the already available technology, in a setup in
which a strong low-frequency and a weak high-frequency laser field
collide head-on with a relativistic nucleus \cite{Piazza2009}. This
pair-production mechanism can be described by the absorption of one
photon from the high-frequency laser field and several additional
photons from the low-frequency laser field \cite{Piazza2009}.

In this article, we study the nonlinear Breit-Wheeler process in the
scenario of a probe photon colliding with two electromagnetic waves.
Pair production off a probe photon colliding with two monochromatic
plane waves and electroweak processes in two monochromatic plane
wave fields were considered for the first time in
Refs.~\cite{Lyulka1974, Lyulka1977, Lyulka1975}. In this article, we
focus on the high multiphoton nonlinear phenomenon of the nonlinear
Breit-Wheeler process in the case of a probe photon $\gamma'$
colliding with a low-frequency and a high-frequency electromagnetic
wave that propagate in the same direction. The results obtained are
compared and contrasted with the results obtained in the nonlinear
Breit-Wheeler process for the case of the probe photon colliding
with one electromagnetic wave only.

The article is organized as follows. In Sec.~\ref{sec:BaF}, we
present the basic formalism of the nonlinear Breit-Wheeler process
in the case of the collision of a probe photon with two
electromagnetic waves. In Sec.~\ref{sec:NA}, we present our detailed
discussions on the basis of the numerical analysis of the high
multiphoton phenomenon of a photon colliding with a low-frequency
and a high-frequency electromagnetic wave that propagate in the same
direction. A summary and some remarks are given in
Sec.~\ref{sec:sum}. We use units of $\hbar = c = 1$ throughout the
article.

\section{Basic formalism for the pair-production probability}
\label{sec:BaF}

In this article, we adopt the vector potential of electromagnetic
fields as a superposition of two monochromatic plane waves $A^{\mu}$
and $B^{\mu}$ with their wave vectors $k$ and $\kappa$
($k\neq\kappa$) :
\begin{equation} \label{vpot}
  V^{\mu} = A^{\mu} (\varphi) + B^{\mu} (\chi),
\end{equation}
with the phases $\varphi = k\cdot x$, $\chi = \kappa\cdot x$. We
consider the case of two waves propagating in the same direction,
satisfying the conditions $k\cdot \kappa = 0$, $k\cdot B = 0$,
$\kappa\cdot A = 0$, and $A\cdot B = 0$. More specifically, we
assume the vector potentials of the two plane waves to be
\begin{eqnarray}
  A^{\mu} = a^{\mu} \cos{\varphi};~k^{\mu} = (\omega_1,0,0,\omega_1);~a^{\mu} =
  |a|(0,1,0,0),~& &\label{apot}\\
  B^{\mu} = b^{\mu} \cos{\chi_b};~\kappa^{\mu} = (\omega_2,0,0,\omega_2);~b^{\mu} =
  |b|(0,0,1,0), & &\label{bpot}
\end{eqnarray}
where $\chi_b \equiv \chi + \delta$, $\delta$ is the phase shift,
and $\omega_1$ and $\omega_2$ are the frequencies of the two plane
waves.

Similar to the calculations of the nonlinear Breit-Wheeler process
(\ref{one-wave}) of a probe photon colliding with an electromagnetic
wave \cite{Nikishov1964, Nikishov1965, Nikishov1967, Narozhnyi1965,
Ritus1985}, we study the probability of $e^{-}e^{+}$ pair production
off a probe photon colliding with two plane waves,
\begin{equation}\label{two-wave}
\gamma' + \bar{n}_1 \gamma_1 + \bar{n}_2 \gamma_2 \rightarrow e^{+}
+ e^{-},
\end{equation}
by using the Volkov solutions of the Dirac equation in two plane
waves to calculate the scattering amplitude in the Furry picture of
QED. In the Furry picture, the $S$ matrix element of the scattering
amplitude in the tree level for this process (\ref{two-wave}) is
expressed as
\begin{equation} \label{sfig}
  S_{fi} = -\mathbf{i} e\int d^4x \bar{\psi}_{p_1\sigma_1}^{e^-}
  \frac{{\rm{e}}^{-\mathbf{i}k'\cdot
  x}}{\sqrt{2k'_0}} \slashed{\varepsilon} \psi_{p_2\sigma_2}^{e^+}, 
\end{equation}
where $\varepsilon$ is the polarization of the probe photon and $k'$
is its momentum. $\psi_{p_1\sigma_1}^{e^-}$ and
$\psi_{p_2\sigma_2}^{e^+}$ are the wave functions of the outgoing
electron and positron in two plane waves. $p_1$ and $p_2$ are the
momenta of the electron and positron and $\sigma_1$ and $\sigma_2$
are their spins, respectively. $\mathbf{i}$ represents the imaginary
unit. We also introduce the invariant parameters of the
electromagnetic fields as $\xi_1 = |e| |a|/m$, $\zeta_1 =(k\cdot k')
\xi_1 / m^2$, $\xi_2 = |e| |b|/m$, and $\zeta_2 = (\kappa \cdot k')
\xi_2/m^2$, with $m$ being the mass of electrons.

The solutions of the Dirac equation in a background plane wave were
found by Volkov in 1935 \cite{Volkov1935, Berestetskii1982}.
Following a similar method, one can obtain the solutions of the
Dirac equation in two background plane waves \cite{Lyulka1974,
Lyulka1977, Lyulka1975, Pardy2006}. In the case of the fields with
the vector potential of Eqs.~(\ref{vpot})-(\ref{bpot}), the solution
of the Dirac equation for the electron is
\begin{eqnarray}
  \psi_{p_1\sigma_1}^{e^-} &=& \left[ 1+\frac{e \slashed{k} \slashed{A}}{2(k\cdot p_1)}
  +\frac{e \slashed{\kappa} \slashed{B}}{2(\kappa\cdot p_1)} \right]
  u_{_{p_1\sigma_1}}  {\rm{e}}^{-\mathbf{i}q_1\cdot x} \nonumber\\
  &\times& \exp{\left[ -\mathbf{i} \frac{e(p_1\cdot a)}{(k\cdot p_1)} \sin{\varphi}
  + \mathbf{i} \frac{e^2 a^2}{8(k\cdot p_1)} \sin{2\varphi} \right]}
  \nonumber\\
  &\times & \exp{\left[ -\mathbf{i} \frac{e(p_1\cdot b)}{(\kappa\cdot p_1)} \sin{\chi_b}
  + \mathbf{i} \frac{e^2 b^2}{8(\kappa\cdot p_1)} \sin{2\chi_b}
  \right]}. \label{phies}
\end{eqnarray}
The wave function of the positron $\psi_{p_2\sigma_2}^{e^+}$ can be
obtained from Eq.~(\ref{phies}) by the substitutions $p_1\rightarrow
p_2$, $q_1\rightarrow q_2$, and $u_{_{p_1\sigma_1}}\rightarrow
v_{_{p_2\sigma_2}}$, where $u_{_{p_1\sigma_1}}$ and
$v_{_{p_2\sigma_2}}$ are, respectively, the spinor of a free
electron and a free positron. The effective momenta and mass are
\begin{eqnarray}
  q_i^{\mu} &=& p_i^{\mu} - \frac{e^2 a^2}{4(k\cdot p_i)} k^{\mu} -
  \frac{e^2 b^2}{4(\kappa \cdot p_i)} \kappa^{\mu}, \quad i =1,2, \label{qdef} \\
  m_{*}^2 &=& m^2 - \frac{e^2 a^2}{2} - \frac{e^2 b^2}{2}.
  \label{mstar}
\end{eqnarray}

In order to calculate the pair-production probability, the following
Fourier series \cite{Ritus1985, Nikishov1964} were introduced
\begin{eqnarray}
  \cos^s{\varphi} {\rm{e}}^{\mathbf{i}(-\alpha_1 \sin{\varphi} +  \beta_1 \sin{2\varphi})}
  = \sum_{n_1=-\infty}^{\infty} \mathcal{A}_s(n_1\alpha_1\beta_1) {\rm{e}}^{-\mathbf{i} n_1
  \varphi},~~~~ & & \label{fouriera}\\
  \cos^s{\chi_b} {\rm{e}}^{\mathbf{i}(-\alpha_2 \sin{\chi_b}
  + \beta_2 \sin{2\chi_b})} = \sum_{n_2=-\infty}^{\infty} \mathcal{B}_s(n_2\alpha_2\beta_2)
  {\rm{e}}^{-\mathbf{i} n_2 \chi_b}, & & \label{fourierb}
\end{eqnarray}
where
\begin{equation} \label{alpha1beta1def}
  \alpha_1 = e\left[\frac{p_2\cdot a}{k\cdot
  p_2} -\frac{p_1\cdot a}{k\cdot p_1}\right];~\beta_1 =
  -\frac{e^2 a^2}{8(k\cdot p_1)} - \frac{e^2 a^2}{8(k\cdot
  p_2)}.
\end{equation}
By the substitutions of $a\rightarrow b$ and $k\rightarrow \kappa$,
$\alpha_2$ and $\beta_2$ can be obtained from $\alpha_1$ and
$\beta_1$ of Eq.~(\ref{alpha1beta1def}), respectively.
$\mathcal{A}_s(n_1\alpha_1\beta_1)$ and
$\mathcal{B}_s(n_2\alpha_2\beta_2)$ are expressed by Bessel
functions, and obey the following relations \cite{Ritus1985,
Nikishov1964}:
\begin{eqnarray}
  (n_1 - 2\beta_1)\mathcal{A}_0 - \alpha_1 \mathcal{A}_1 + 4 \beta_1 \mathcal{A}_2
  &=& 0, \label{relationa}\\
  (n_2 - 2\beta_2)\mathcal{B}_0 - \alpha_2 \mathcal{B}_1 + 4 \beta_2 \mathcal{B}_2
  &=& 0, \label{relationb}
\end{eqnarray}
where we introduce the notations $\mathcal{A}_s := \mathcal{A}_s(n_1
\alpha_1 \beta_1)$ and $\mathcal{B}_s := \mathcal{B}_s(n_2 \alpha_2
\beta_2)$.

Following the approach presented in Ref.~\cite{Ritus1985}, with the
help of Eqs.~(\ref{phies})-(\ref{relationb}), we calculate the
pair-production probability ($W$) per unit volume and per unit time
for the nonlinear Breit-Wheeler process (\ref{two-wave}) of pair
production off a probe photon colliding with the two plane waves
with the vector potential of Eqs.~(\ref{vpot})-(\ref{bpot}) by
squaring the $S$ matrix element (\ref{sfig}), summing over the
polarizations of the outgoing electron and positron, averaging over
the polarizations of the probe photon, and integrating over the
final states $d^3 q_1 d^3 q_2 (2\pi)^{-6}$ of the positron and the
electron,
\begin{equation} \label{ppptv}
  W = \frac{e^2 n_{\gamma}}{16\pi^2 k'_0} \sum_{n_1,n_2} \int_{0}^{2\pi} d\phi
  \int_1^{u_s}
  \frac{du}{u\sqrt{u(u-1)}}\mathcal{M}(n_1,n_2),
\end{equation}
where $u = (k\cdot k')^2/4(k\cdot q_1)(k\cdot q_2)$ and $\phi$ is
the angle between the ($\vec{k}$, $\vec{q}_1$) and ($\vec{k}$,
$\vec{a}$) planes in a system in which $\vec{k}$ and $\vec{k'}$ are
oppositely directed. In the probability (\ref{ppptv}), $n_{\gamma}$
is the average density of the probe photon beam and
$\mathcal{M}(n_1,n_2)$ is given by
\begin{eqnarray}
  \mathcal{M}(n_1,n_2) = m^2 \mathcal{A}_0^2 \mathcal{B}_0^2~~~~~~~~~~~~~~~~~~~~~~
  ~~~~~~~~~~~~~~~~~~~~~~~~~~~~~~~& & \nonumber\\
  + e^2 a^2 \left[1 - \frac{(k\cdot k')^2}{2(k\cdot p_1) (k\cdot
  p_2)}\right] (\mathcal{A}_1^2 \mathcal{B}_0^2 - \mathcal{A}_0 \mathcal{A}_2
  \mathcal{B}_0^2)~~~~& & \nonumber \\
  + e^2 b^2 \left[1 - \frac{(\kappa\cdot k')^2}{2(\kappa\cdot p_1) (\kappa\cdot
  p_2)}\right] (\mathcal{A}_0^2 \mathcal{B}_1^2 - \mathcal{A}_0^2
  \mathcal{B}_0 \mathcal{B}_2). & & \label{mtwo}
\end{eqnarray}
In addition, the maximum value for $u$ in Eq.~(\ref{ppptv}) is
\begin{equation} \label{umax}
  u_s = \frac{n_1(k\cdot k') + n_2 (\kappa\cdot k')}{2m_{*}^2}.
\end{equation}

The positive values of $n_1$ ($n_2$) physically indicate that $n_1$
($n_2$) photons are absorbed from the first wave (\ref{apot}) [the
second wave (\ref{bpot})] in the process, while the negative values
of $n_1$ ($n_2$) physically indicate $|n_1|$ ($|n_2|$) photons are
emitted into the first wave (the second wave) in the process. The
numbers $n_1$ and $n_2$ of photons absorbed from (emitted into) the
waves must satisfy the threshold condition
\begin{equation} \label{thresholdv}
  n_1(k\cdot k') + n_2 (\kappa\cdot k') > 2m_{*}^2.
\end{equation}
The summation of $n_1$ and $n_2$ in Eq.~(\ref{ppptv}) must satisfy
the condition (\ref{thresholdv}). Each $n_1$ and $n_2$ in the
probability (\ref{ppptv}) corresponds to a four-quasimomentum
conservation law
\begin{equation} \label{emcl}
  n_1 k + n_2 \kappa + k' = q_1 + q_2.
\end{equation}
As shown in the threshold condition (\ref{thresholdv}), negative
values of $n_1$ or $n_2$ are allowed in the nonlinear Breit-Wheeler
process of pair production off a probe photon colliding with two
monochromatic plane waves. This shows the possibility of
Breit-Wheeler pair production by the absorption of photons from one
of these two waves and the probe photon with simultaneous photon
emission into the other wave. This is quite different from the
nonlinear Breit-Wheeler process of pair production off a probe
photon colliding with one plane wave only, for which negative values
of $n_1$ ($n_2$) are not allowed by the threshold condition.

When one of the two plane waves is absent, the pair-production
probability (\ref{ppptv}) reduces to the result of the nonlinear
Breit-Wheeler process (\ref{one-wave}) in one plane wave obtained by
Refs.~\cite{Ritus1985, Nikishov1964}.  More specifically, for the
case of $|b|=0$, we obtain the pair-production probability
\begin{equation} \label{pppov}
  W = \frac{e^2 m^2 n_{\gamma}}{16\pi^2 k'_0} \sum_{n_1>n_0} \int_{0}^{2\pi} d\phi
  \int_1^{u_{s_1}}
  \frac{du}{u\sqrt{u(u-1)}}\mathcal{M}_a(n_1),
\end{equation}
with
\begin{equation} \label{monea}
  \mathcal{M}_a(n_1) = m^2 \mathcal{A}_0^2 + e^2 a^2
  \left[1 - \frac{(k\cdot k')^2}{2(k\cdot p_1) (k\cdot
  p_2)}\right] (\mathcal{A}_1^2- \mathcal{A}_0 \mathcal{A}_2),
\end{equation}
the threshold value $n_0 = 2m_{*}^2/(k\cdot k')$, and the maximum
value $u_{s_1} = n_1(k\cdot k')/2m_{*}^2$. The effective mass
$m_{*}$ in this case is given by Eq.~(\ref{mstar}) by setting $b^2
=0$. Using formulas (\ref{pppov})-(\ref{monea}),
Refs.~\cite{Ritus1985, Nikishov1964} studied pair production off a
probe photon colliding with a monochromatic plane wave in the cases
of $\xi_1 \gg 1$ and $\xi_1\ll 1$. Their results showed that (i) in
the case $\xi_1\ll 1$, the process yields to the perturbation
process as considered first by Breit and Wheeler; (ii) in the case
$\xi_1\gg 1$, the process yields to the case of a constant crossed
field \cite{Ritus1985, Nikishov1964}.

After having obtained the probability (\ref{ppptv})-(\ref{mtwo}) of
pair production off a probe photon colliding with two monochromatic
plane waves, we occasionally find a similar study in
Ref.~\cite{Lyulka1974}, where the pair-production probability was
numerically calculated for the case in which the invariant
parameters $\xi_1$, $\zeta_1$, $\xi_2$, and $\zeta_2$ are of the
order of unity. In this article we use
Eqs.~(\ref{ppptv})-(\ref{mtwo}) to present an analysis of the
phenomenon of pair production with simultaneous photon emission into
the low-frequency wave and high multiphoton phenomenon in this
process (\ref{two-wave}) in the case of $\omega_2 \gg \omega_1$. We
select the parameters of the electromagnetic fields and the probe
photon close to the values used in the SLAC-E-144 experiment
\cite{Burke1997, Bamber1999}. We calculate the pair-production
probability and the spectra of the created pair in this process. For
the purpose of necessary comparisons, we also present the same
analysis of the one plane-wave process (\ref{one-wave}) (i.e., one
of the two plane waves is absent). As a result, we quantitatively
compare and contrast the multiphoton phenomenon in the two
plane-wave process (\ref{two-wave}) and one plane-wave process
(\ref{one-wave}).

It is important to point out that, to calculate the squared
$|S_{fi}|^2$ of the $S$-matrix element (\ref{sfig}), one has to
perform the double sum over $n'_1$ and $n'_2$ in addition to the
double sum over $n_1$ and $n_2$ satisfying the relation
\begin{equation} \label{intfp}
  n_1 k + n_2 \kappa = n'_1 k + n'_2 \kappa,
\end{equation}
since $n_1$ and $n_2$ denote the modes of the Fourier series
(\ref{fouriera}) and (\ref{fourierb}) of the $S$-matrix element
$S_{fi}$ (\ref{sfig}), whereas $n'_1$ and $n'_2$ denote the modes of
the Fourier series (\ref{fouriera}) and (\ref{fourierb}) of the
complex conjugate of the $S$-matrix element $S_{fi}^{\dag}$. If the
frequencies of the two waves are commensurate (the ratio
$\omega_2/\omega_1$ is a rational number), the quantum interferences
of the amplitudes corresponding to $n'_1\not=n_1$, $n'_2\not=n_2$,
and satisfying the relation (\ref{intfp}), arise
\cite{Narozhny2000}. The pair-production probability ($\sim
|S_{fi}|^2$) receives contributions from the quantum interferences
of the amplitudes, in which there is a phase factor
$\exp{[i(n'_2-n_2)\delta]}$ for the given ($n'_2$, $n_2$). As shown
in Ref.~\cite{Narozhny2000}, the optimal value of the frequency
ratio $\omega_2/\omega_1$ for observing interference effects is $3$,
i.e., $\omega_2=3~\omega_1$. If the frequencies of the two waves are
incommensurate, there are no solutions of $n'_1\not=n_1$ and
$n'_2\not=n_2$ in Eq.~(\ref{intfp}); hence, the contributions from
the quantum interferences of the amplitudes $n'_1\not=n_1$ and
$n'_2\not=n_2$ to the $|S_{fi}|^2$ vanish, Eqs.~(\ref{ppptv}) and
(\ref{mtwo}) obtained from the amplitudes $n'_1=n_1$ and $n'_2=n_2$
are exact results \cite{Lyulka1974, Narozhny2000}, and the
dependence of the pair-production probability on the phase shift
$\delta$ vanishes \cite{Narozhny2000}.

In this article, in order to analyze the phenomenon of pair
production with simultaneous photon emission into the low-frequency
wave and high multiphoton phenomenon in this two-wave process
(\ref{two-wave}) in the case of $\omega_2 \gg \omega_1$, we select
the frequencies of these two waves to be $\omega_2=100~\omega_1$
(see Sec.~\ref{sec:parameter} for details). Although the frequencies
of these two waves are commensurate, we approximately adopt the
exact pair-production probability (\ref{ppptv}) for $n'_1=n_1$ and
$n'_2=n_2$ to study pair production in our case by ignoring the
contributions from the quantum interferences of the amplitudes. The
reasons are given as follows. In the case $\omega_2 \gg \omega_1$,
the effects of quantum interferences are present only in the very
large wave modes ($n_1\not=n_1'$ and $|n_1|$ and/or $|n_1'|\gg 1$)
\cite{Narozhny2000}. The contributions of the processes with large
wave modes ($n_1\not=n_1'$ and $|n_1|$ and/or $|n_1'|\gg 1$) are
suppressed by the weak $\xi_1$ and $\xi_2$ considered in this
article. Therefore, the effects of quantum interferences are
expected to be small in our case. Nevertheless, in numerical
calculations we have made some self-consistency checks to make sure
that the interference effects are indeed negligible in our case for
studying the contributions of multiphoton processes to the
pair-production rate. However, we would like to mention that we are
interested in investigating the interference effects in future
studies.

\section{Numerical analysis}
\label{sec:NA}

\subsection{Plane-wave and probe photon fields}
\label{sec:parameter}

Based on the high-energy photons and technology for laser beams used
in the SLAC-E-144 experiment, we consider the following parameters
for high-energy photons and electromagnetic plane waves. The
intensities of the electromagnetic fields are selected to be $I_1 =
I_2 = 10^{18}~\rm{W/cm^2}$. The intensity $I_i$ ($i=1,2$) of the
electromagnetic field is related to the field strength parameter
$\xi_i$ by
\begin{equation} \label{relationxiI}
\xi_i^2 \approx 1.13\times 10^{-18}\,[\omega_i({\rm{eV}})]^{-2}\,
I_i({\rm{W/cm^2}}).
\end{equation}
For an electromagnetic field with intensity $I =
10^{18}~\rm{W/cm^2}$ and frequency $\omega \sim 1~{\rm{eV}}$ (the
optical regime), the field strength parameter $\xi$ is of the order
of unity. Also we set $\omega_2 = 100~\omega_1$ throughout the
following calculations.

One of the purposes in this article is to study the difference
between the nonlinear Breit-Wheeler process of pair production off a
probe photon colliding with two plane waves and the nonlinear
Breit-Wheeler process of pair production off a probe photon
colliding with one plane wave. For the sake of comparison, we choose
the energy of the probe photon so that the probability of pair
production off the probe photon colliding with one of these two
plane waves is almost the same as the probability of pair production
off the probe photon colliding with the other plane wave, in the
regime of the one used in the SLAC-E-144 experiment ($\sim 30~
{\rm{GeV}}$). This means that for each value of $\xi_1$ ($\omega_1$)
we have a corresponding value of $k'_0$ (see Table \ref{ppenergy}).

\begin{table}[h]  \addtolength{\tabcolsep}{8pt}
\begin{center}
\begin{tabular}{l|c|c|c|c|c}
\hline  $\omega_1$ (eV) & $2.5$ & $3$    & $3.5$  & $4$    & $4.5$ \\
\hline  $k'_0$ (GeV)    & $38$  & $31.4$ & $27.4$ & $26.4$ & $29.8$
\\
\hline
\end{tabular}
\end{center}
\caption{The energy $k'_0$ of the probe photon for each selected
value of the frequency $\omega_1$.} \label{ppenergy}
\end{table}

It is worth mentioning the recent article \cite{Jansen2013} that
presents the analysis of scalar pair production off a high-energy
photon colliding with a bifrequent laser wave within the framework
of laser-dressed scalar QED. Using the parameters of the
electromagnetic fields of two laser beams $\omega_2 \sim m$,
$\omega_1 \sim 0.1~\omega_2$, $\xi_1 \sim 1$, and $\xi_2 \sim
10^{-2}$, and the energy of the probe photon $k_0'\sim m$,
Ref.~\cite{Jansen2013} shows that the pair-production rate can be
largely enhanced compared with the case in the absence of the
high-frequency laser wave. In the present article, we actually
perform our analysis in the framework of laser-dressed spinor QED in
order to take into account all contributions from laser-dressed
spinor wave functions (\ref{phies}) of electrons and positrons. The
effect of enhancement \cite{Jansen2013} is also observed in our
analysis with the different parameters of laser beams and probe
photons. However, in this article, we present (and focus on) the
phenomenon of pair production with simultaneous photon emission into
the low-frequency wave and high multiphoton nonlinear phenomenon in
the nonlinear Breit-Wheeler process (\ref{two-wave}) of pair
production off a probe photon colliding with a low-frequency wave
and a high-frequency wave. We calculate the pair-production
probability and the spectra of the created pair, and compare and
contrast these results with the results obtained in the case of the
probe photon colliding with each of these two plane waves, to
provide a possible way to access the phenomenon of photon emission
into the wave and the high multiphoton phenomenon. Besides, we
purposely select the energy of the probe photon and parameters of
the electromagnetic fields of two laser beams on the basis of the
SLAC-E-144 experiment, so as to closely relate our results to the
experimental situation.

\begin{figure}[h]
\includegraphics[width=0.98\columnwidth]{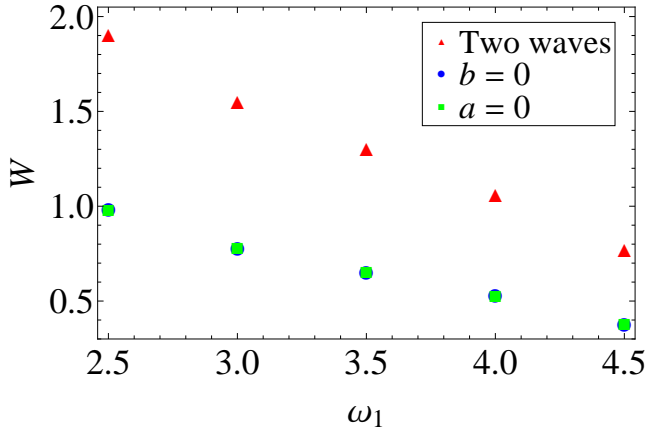}
\caption{(Color online) The pair-production probability $W$ is
calculated for selected values of the frequency $\omega_1$. We
normalize the probability $W$ by $10^{-6} n_{\gamma}~{\rm{eV}}$, the
average density $n_\gamma$ of the probe photon beam is in units of
$({\rm eV})^3$, and the frequency $\omega_1$ is in units of eV. We
show the results in the following cases: (i) a probe photon
colliding with two plane waves; (ii) the probe photon colliding with
each of these two plane waves. Here we choose the frequencies
$\omega_2 = 100~\omega_1$ and the intensities of the fields $I_1 =
I_2 = 10^{18}~\rm{W/cm^2}$. For each value of $\omega_1$, the energy
of the corresponding probe photon ($\sim 30$GeV) is selected (see
Table \ref{ppenergy}) to make the probability of pair production off
the probe photon colliding with one of the two plane waves and the
probability of pair production off the probe photon colliding with
the other plane wave to be almost the same (see the overlapping of
full circles and squares).} \label{wpplot}
\end{figure}

\subsection{Two and one plane-wave cases}

In Fig.~\ref{wpplot}, we show the pair-production probability $W$ in
the case of a probe photon colliding with two plane waves for
different values of the frequency $\omega_1$ and the pair-production
probability $W$ in the case of the probe photon colliding with each
of these two plane waves. As shown in Fig.~\ref{wpplot}, the
pair-production probability (we denote here as $W_t$) in the case of
a probe photon colliding with two plane waves is larger than its
corresponding probabilities ($W_a$ and $W_b$, $W_a\approx W_b$) in
the case of the probe photon colliding with each of these two plane
waves, but slightly smaller than the sum of these two probabilities
($W_a+ W_b$), i.e.,
\begin{equation} \label{wtwovsone}
  (W_a+W_b) \gtrapprox W_t > W_{a,b}.
\end{equation}
Here we introduce the notations $W_a=W|_{|b|=0}$ and
$W_b=W|_{|a|=0}$ for the probability of pair creation off the probe
photon colliding with each of these two plane waves, respectively,
which are calculated by Eqs.~(\ref{pppov})-(\ref{monea}). It is
necessary to clarify here that the relation (\ref{wtwovsone}) is not
true, in general, but only holds for the parameters of the laser
waves and probe photon considered in this article; e.g., it does not
hold for the case discussed in Ref.~\cite{Jansen2013}. The result
(\ref{wtwovsone}) can be understood from the point that the
effective masses (\ref{mstar}) of electrons and positrons in the two
plane waves are slightly larger than the effective masses of
electrons and positrons in one of these two plane waves. Therefore,
in the case of two plane waves, the contribution to the
pair-production probability from each plane wave is slightly
suppressed by the slightly large effective mass. This leads to the
result that $W_t\lessapprox (W_a+W_b)$. We will give some further
discussions below.

\begin{figure}[h]
\includegraphics[width=0.98\columnwidth]{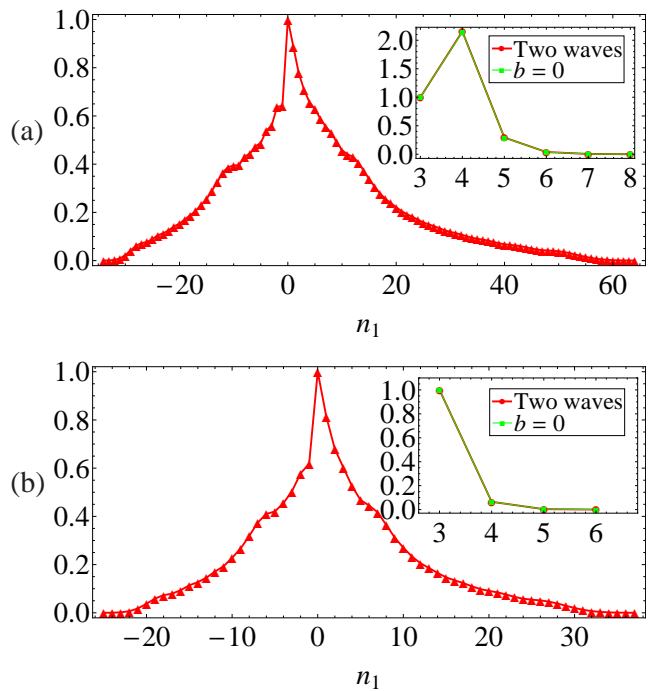}
\caption{(Color online) The normalized pair-production probability
$W_1(n_1)$ is plotted as a function of $n_1$. The normalized
pair-production probability $W_{ar}(n_1)$ for the case of one plane
wave ($|b|=0$) and $W_0(n_1)$ are plotted in the inset. (a)
$\omega_1 = 2.5~{\rm{eV}}$ and (b) $\omega_1 = 4~{\rm{eV}}$. Here we
use the same parameters of the plane waves and probe photons as the
ones used in Fig.~\ref{wpplot}.} \label{norder}
\end{figure}

\subsection{The high multiphoton phenomenon and pair production with
simultaneous photon emission}

In order to the phenomenon of pair production with simultaneous
photon emission into the low-frequency wave and high multiphoton
phenomenon in this nonlinear Breit-Wheeler process (\ref{two-wave}),
we study the pair-production probability for the process with the
given $n_1$ and $n_2$ related to the given numbers ($|n_1|, |n_2|$)
of photons absorbed from (emitted into) the two plane waves,
\begin{equation} \label{ppptvn}
  W_t(n_1,n_2) = \frac{e^2 n_{\gamma}}{16\pi^2 k'_0}\int_{0}^{2\pi}
  d\phi
  \int_1^{u_s} \frac{du}{u\sqrt{u(u-1)}}\mathcal{M}(n_1,n_2).
\end{equation}
For the following discussions, we define the normalized
pair-production probability for the process with the given $n_1$ and
$n_2$ as
\begin{equation} \label{ppptvnr}
  W_{n_2}(n_1) = W_t(n_1,n_2)/W_t(n^{\rm thre}_1(n_2),n_2),
\end{equation}
where $n^{\rm thre}_1(n_2)$ is the minimum value of $|n_1|$
determined by Eq.~(\ref{thresholdv}) for a fixed value of $n_2$. For
a comparison with the case of one plane wave ($|b|=0$), analogously
to Eq.~(\ref{ppptvnr}), we also define the normalized
pair-production probability for the process with a given $n_1$ in
the case when the high-frequency wave is absent ($|b|=0$) as
\begin{equation} \label{pppovnr}
  W_{ar}(n_1) = W_a(n_1)/W_a(n_{0}),
\end{equation}
where $W_a(n_1)$ is obtained from Eq.~(\ref{pppov}),
\begin{equation} \label{pppovn}
  W_a(n_1) = \frac{e^2 n_{\gamma}}{16\pi^2 k'_0}\int_{0}^{2\pi} d\phi
  \int_1^{u_{s_1}} \frac{du}{u\sqrt{u(u-1)}}\mathcal{M}_a(n_1).
\end{equation}

The results of $W_0(n_1)$, $W_1(n_1)$, and $W_{ar}(n_1)$ for the
cases of $\omega_1 = 2.5~{\rm{eV}}$ and $4~{\rm{eV}}$ are shown in
Fig.~\ref{norder}. We show these results only for the two lowest
numbers ($n_2 = 0,1$) of photons absorbed from the high-frequency
wave field. The contributions from $n_2 > 1$ are very small because
the $\xi_2$ under consideration is very small ($\xi_2 \ll 1$). This
is in agreement with the result of Refs.~\cite{Ritus1985,
Nikishov1964}. In addition, the contributions from negative values
of $n_2$ are also very small. The reasons as follows. According to
the threshold condition (\ref{thresholdv}), when $n_2$ is negative,
there must be a very large number of photons absorbed from the
low-frequency wave ($n_1 \gg 1$). However, the contributions from
the processes with large $n_1$ ($n_1 \gg 1$) are suppressed by the
weak $\xi_1$ considered in this article.

It is shown in Fig.~\ref{norder} that $W_0(n_1)$ is rather close to
$W_{ar}(n_1)$ for each selected value of $\omega_1$. This can be
understood as follows. In our calculations for $\xi_1\gg \xi_2$, the
effects of the high-frequency wave on the wave function of electrons
and positrons such as the spin correction, phase correction, and
effective mass are small, compared to the effects of the
low-frequency wave. In this regard, the influence of the
high-frequency wave on the pair is very small for the process in
which there is no photon absorbed from the high-frequency wave. More
explicitly, this can be seen from the expression of $\mathcal{B}_s$
in Refs.~\cite{Ritus1985, Nikishov1964} and the properties of the
Bessel functions (see, e.g., Ref.~\cite{Abramowitz1965}) that
$\mathcal{B}_0|_{n_2 =0} \sim~1$, $\mathcal{B}_1|_{n_2 = 0} \sim~0$,
and $\mathcal{B}_2|_{n_2=0} \sim~1/2$, for the parameters of the
plane waves we use here. As a result, it is shown from
Eqs.~(\ref{ppptvn}) and (\ref{pppovn}) that $W_0(n_1)$ is close to
$W_{ar}(n_1)$.

In Fig.~\ref{norder}, $W_{ar}(n_1)$ decreases rapidly as the value
of $n_1$ increases. This result is in agreement with the discussion
in Refs.~\cite{Ritus1985, Nikishov1964} for the field strength
parameter $\xi_1<1$. In this case, the main contributions to the
probability of pair production off a probe photon colliding with one
monochromatic plane wave are from the processes in which the number
of photons absorbed from the wave is near the threshold value when
the field strength parameter $\xi_1<1$. On the contrary, $W_1(n_1)$
shows the high multiphoton phenomenon. As shown in
Fig.~\ref{norder}, $W_1(n_1)$ has a significant value even when the
absorbed photon number $n_1$ is around $20$. The absorption of one
photon from the high-frequency wave enhances the multiphoton
processes of the low-frequency wave. It is also shown in
Fig.~\ref{norder} that when increasing the frequency $\omega_1$, the
multiphoton phenomenon of $W_1(n_1)$ is suppressed, consistent with
the discussions \cite{Ritus1985, Nikishov1964} in which the
multiphoton phenomenon dominantly depends on the field strength
parameters $\xi_1$ and $\xi_2$. When the frequencies increase and
the intensities do not change, $\xi_1$ and $\xi_2$ decrease, leading
to a lower multiphoton phenomenon.

In Fig.~\ref{norder}, we show the phenomenon of Breit-Wheeler pair
production with simultaneous photon emission (represented by the
negative values of $n_1$) into the low-frequency wave by absorbing
one photon from the high-frequency wave and the probe photon. In
Fig.~\ref{norder}, the high multiphoton phenomenon is also shown in
this Breit-Wheeler process with simultaneous photon emission into
the low-frequency wave since $W_1(-|n_1|)$ has a significant value
even when $|n_1|$ is large (e.g.~$|n_1|\sim 20$). In addition,
$W_1(-|n_1|)$ is smaller than $W_1(|n_1|)$ for a fixed $|n_1|$. This
implies that, absorbing one photon from the high-frequency wave and
the probe photon, the probability of Breit-Wheeler pair production
with photon absorption from the low-frequency wave is larger than
the probability of Breit-Wheeler pair production with simultaneous
photon emission into the low-frequency wave. One of the reasons is
that the former has a larger phase space of Breit-Wheeler pair
production than the latter. We stress once again that this
phenomenon of pair production with simultaneous photon emission into
the wave cannot happen in the nonlinear Breit-Wheeler process of
pair production off a probe photon colliding with one plane wave
only (see the inset in Fig.~\ref{norder}).

In addition, we want to mention that, as shown in
Fig.~\ref{norder}(a) for the case $\omega_1=2.5~{\rm{eV}}$,
$W_{ar}(n_1)$ first increases and then decreases, as $n_1$
increases. This is mainly because the selected value of $\zeta_1$ in
this case leads to a threshold value $n_0$ very close to $3$. As a
result, the phase space in the integration of Eq.~(\ref{pppovn}) for
the case $n_1 = 3$ is rather small compared to the one of the case
$n_1 = 4$. This explains the results of $n_1 = 3$ and $n_1 = 4$
shown in Fig.~\ref{norder}(a).

\begin{figure}[h]
\includegraphics[width=0.9\columnwidth]{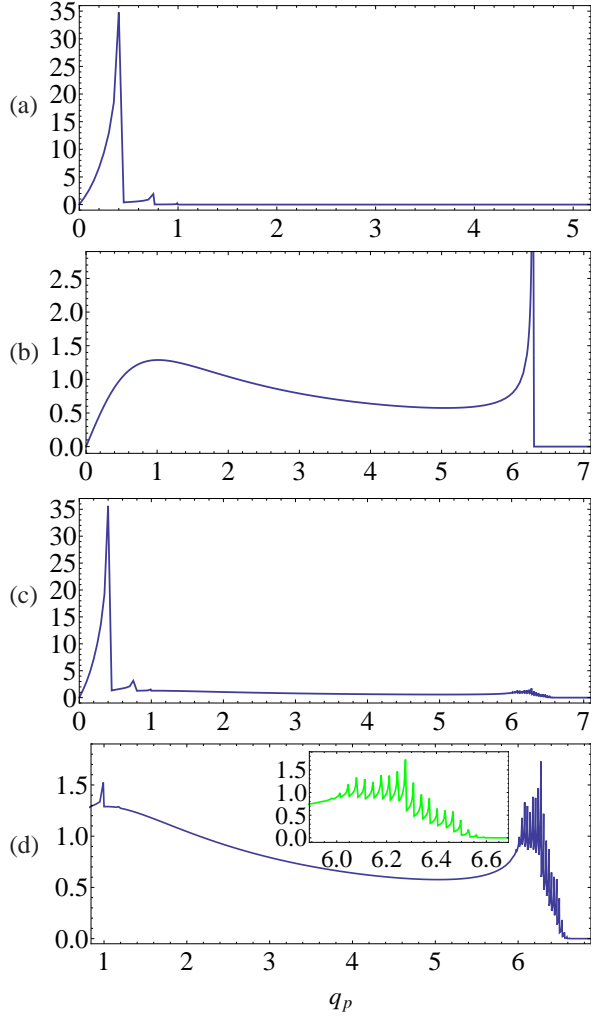}
\caption{(Color online) The differential pair-production probability
$d W/d q_p$ is normalized by $10^{-7} n_{\gamma}~{\rm{eV}}$. We plot
it in terms of the transverse component $q_p$. Here we use the same
parameters of the plane waves and probe photons adopted in
Fig.~\ref{wpplot} for $\omega_1=4~{\rm{eV}}$. (a): $d W/d q_p$ for
the one-wave case of $|b|=0$. (b): $d W/d q_p$ for the one-wave case
of $|a|=0$. (c) $d W/d q_p$ for the case of two waves. (d) Detailed
results of $d W/d q_p$ for the case of two waves.} \label{spectra40}
\end{figure}

\subsection{The pair spectrum}

Now we turn to the spectra of the $e^-~e^+$ pair to give a further
understanding and consequence of the phenomenon of pair production
with simultaneous photon emission into the low-frequency wave and
high multiphoton phenomenon shown in Fig.~\ref{norder}. From the
four-quasimomentum conservation law (\ref{emcl}) and the definition
of the invariant variable $u$, we obtain the transverse component
$|q_{\bot}|$ of $\vec{q}_1$ and $\vec{q}_2$, which are perpendicular
to $\vec{k}$ ( $\vec{k}$ and $\vec{k'}$ being oppositely directed),
\begin{equation} \label{qper}
  |q_{\bot}| = m_{*} \sqrt{\frac{u_s}{u} - 1}.
\end{equation}
Introducing the dimensionless transverse momentum $q_p \equiv
|q_{\bot}|/m$, from Eq.~(\ref{ppptv}) we obtain the differential
pair-production probability,
\begin{equation} \label{dppptw}
  \frac{d W}{d q_p} = \sum_{n_1,n_2} \frac{d W_{n_1n_2}}{d q_p},
\end{equation}
and the differential pair-production probability of the process with
the given $n_1$ and $n_2$,
\begin{equation} \label{dppptwn}
  \frac{d W_{n_1n_2}}{d q_p} = \frac{2u q_p}{(m_{*}/m)^2 u_s
  \sqrt{u(u-1)}} \int_{0}^{2\pi} d\phi \mathcal{M}(n_1,n_2).
\end{equation}
From Eq.~(\ref{pppov}), we also obtain the differential
pair-production probability for the case in which the high-frequency
wave is absent ($|b|=0$),
\begin{equation} \label{dpppow}
  \frac{d W}{d q_p} = \sum_{n_1} \frac{d W_a(n_1)}{d q_p},
\end{equation}
where the differential pair-production probability $d W_a(n_1)/d
q_p$ of a given $n_1$ can be obtained from Eq.~(\ref{dppptwn}) by
$\mathcal{M}(n_1,n_2)\rightarrow \mathcal{M}_a(n_1)$ and
$u_s\rightarrow u_{s_1}$ of Eqs.~(\ref{mtwo}) and (\ref{umax}). In
the case when the low-frequency plane wave is absent ($|a|=0$), the
differential pair-production probability $d W/d q_p$ can be obtained
from Eqs.~(\ref{monea}) and (\ref{dpppow}) by the substitutions
$a\rightarrow b$, $k\rightarrow\kappa$, $n_1\rightarrow n_2$, and
$\mathcal{A}_s \rightarrow \mathcal{B}_s$.

\begin{figure}[h]
\includegraphics[width=0.98\columnwidth]{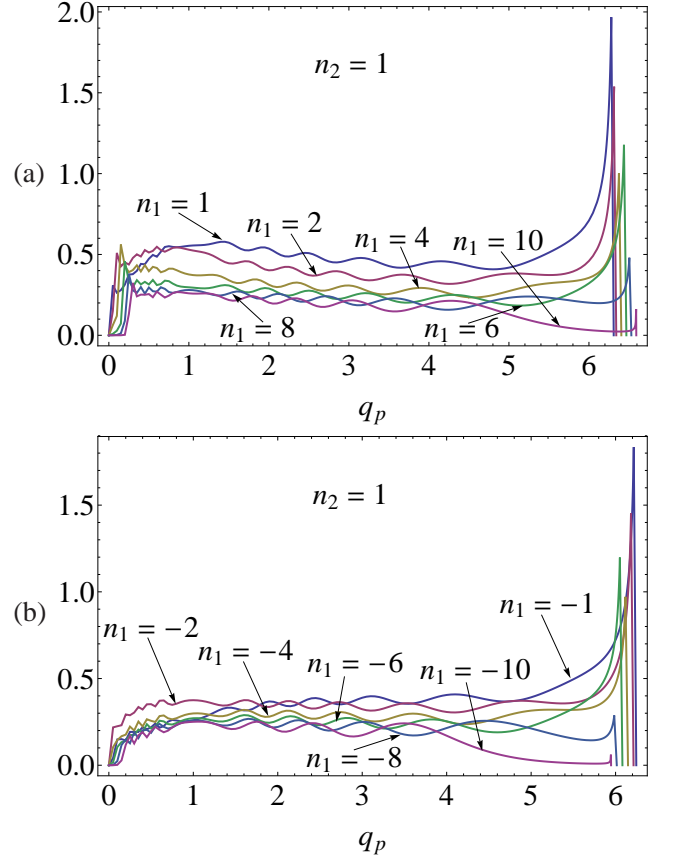}
\caption{(Color online) The differential pair-production probability
$d W_{n_1n_2}/d q_p$ for selected values of $n_1$ and $n_2$ is
normalized by $10^{-8} n_{\gamma}~{\rm{eV}}$. We plot it in terms of
$q_p$ for selecting values (a) $n_1=1,2,4,6,8,10$ with $n_2=1$ and
(b) $n_1=-1,-2,-4,-6,-8,-10$ with $n_2=1$. Here we use the same
parameters of the plane waves and probe photons adopted in
Fig.~\ref{wpplot} for $\omega_1=4~{\rm{eV}}$.} \label{spectrum40s}
\end{figure}

Figure \ref{spectra40} shows the differential pair-production
probability $d W/d q_p$ for both one-wave and two-wave cases.
Equation (\ref{qper}) indicates that the transverse component $q_p$
has a maximal value $q_{p}^{max}$ for the process with the given
$n_1$ and $n_2$ related to the numbers ($|n_1|, |n_2|$) of photons
absorbed from (emitted into) the waves,
\begin{equation} \label{qpmaxtwo}
  q_p^{max} (n_1,n_2)= \frac{m_{*}}{m} \sqrt{u_s - 1}
\end{equation}
for the two-wave case and
\begin{equation} \label{qpmaxa}
  q_p^{max} (n_1)= \frac{m_{*}}{m} \sqrt{u_{s_1} - 1}
\end{equation}
for the case when the high-frequency wave is absent ($|b|=0$). The
maximal value $q_{p}^{max} (n_2)$ of transverse component $q_p$ for
the case in which the low-frequency plane wave is absent ($|a|=0$)
can be obtained from Eq.~(\ref{qpmaxa}) by the substitutions
$a\rightarrow b$, $k\rightarrow\kappa$, and $n_1\rightarrow n_2$.
The existence of $q_{p}^{max}$ is due to the four-quasimomentum
conservation (\ref{emcl}). In Fig.~\ref{spectra40}, we show that,
corresponding to the contributions from the processes with the
different numbers of photons absorbed (emitted), the probability $d
W/d q_p$ has different peaks located around $q_p\sim q_p^{max}$, and
sharply decreases at $q_{p}^{max}$.

As shown in Fig.~\ref{spectra40}(a) for the presence of the
low-frequency wave only, the differential pair-production
probability practically vanishes after $n_1 = 4$, which corresponds
to the maximal value $q_{p}^{max}(n_1)|_{n_1=4}\approx 0.8$. Due to
the fact that the field strength parameter $\xi_2\ll 1$, only the
lowest order ($n_2 = 1$) term has a significant contribution to the
pair-production probability for the case when the low-frequency wave
field is absent ($|a|=0$), as we have discussed above in
Fig.~\ref{norder}. The pair creation process in this case yields to
the perturbation process considered by Breit and Wheeler
\cite{Breit1934}. Its consequent result on the differential
pair-production probability is clearly shown in
Fig.~\ref{spectra40}(b): When $q_p$ is larger than the maximal value
$q_{p}^{max}(n_2)|_{n_2=1}\approx 6.25$ for $n_2 = 1$, the
differential probability sharply decreases to zero.

Compared with the results presented in Figs.~\ref{spectra40} (a) and
\ref{spectra40}(b) for the case of one wave, the multipeak structure
of the differential pair-production probability $d W/d q_p$
presented in Figs.~\ref{spectra40} (c) and \ref{spectra40}(d) for
the case of two waves clearly shows the phenomenon of pair
production with simultaneous photon emission into the low-frequency
wave and the high multiphoton phenomenon. As shown in the inset of
Fig.~\ref{spectra40} (d), the multipeaks of $d W/d q_p$ located at
different values $q_p^{max} (n_1,n_2)$ indicate the multiphoton
processes in which photons of different numbers ($|n_1|$) are
absorbed from (emitted into) the low-frequency wave, in addition to
the absorption of one photon $(n_2=1)$ from the high-frequency wave.
In Fig.~\ref{spectrum40s} (a), we plot $d W_{n_1n_2}/d q_p$, the
detailed spectra of the created $e^-~e^+$ pair, absorbing $n_1$
photons from the low-frequency wave and one additional photon
$(n_2=1)$ from the high-frequency wave. Due to the limit of the
plotting scale, we only show the results of $n_1 = 1,2,4,6,8,10$ in
Fig.~\ref{spectrum40s}(a). In Fig.~\ref{spectrum40s} (b), we plot
the detailed spectra $d W_{n_1n_2}/d q_p$ of the created $e^-~e^+$
pair by absorbing one photon $(n_2=1)$ from the high-frequency wave
and emitting $|n_1|$ photons into the low-frequency wave.

We can learn from Fig.~\ref{spectrum40s} how the multiphoton
(absorption or emission) processes contribute to the total
differential pair-production probability $d W/d q_p$ in
Fig.~\ref{spectra40}. The multipeak structure of the total
differential pair-production probability $d W/d q_p$ for $q_p>6.25$
presented in the inset of Fig.~\ref{spectra40} (d) is due to the
different peaks of the pair spectra after $q_p>6.25$ shown in
Fig.~\ref{spectrum40s} (a), which correspond to the contributions
from the processes with the absorption of photons of different
numbers ($n_1$) from the low-frequency wave, in addition to the
absorption of one photon from the high-frequency wave. The multipeak
structure of the total differential pair-production probability $d
W/d q_p$ for $q_p<6.25$ presented in the inset of
Fig.~\ref{spectra40} (d) is due to the different peaks of the pair
spectra after $q_p>5.8$ presented in Fig.~\ref{spectrum40s} (b),
which correspond to the contributions from the processes with the
emission of photons of different numbers ($|n_1|$) into the
low-frequency wave, in addition to the absorption of one photon from
the high-frequency wave. In addition, the smooth oscillating
structure of $d W_{n_1n_2}/d q_p$ for $q_p\lesssim 5.8$ in
Fig.~\ref{spectrum40s} indicates the interference effect of the two
waves on the phase of the wave function of the pair (\ref{phies}).
The results presented in Figs.~\ref{spectra40} and \ref{spectrum40s}
provide a possible way to access the phenomenon of pair production
with simultaneous photon emission into the low-frequency wave and
high multiphoton (absorption and emission) phenomenon in the
nonlinear Breit-Wheeler process of pair production off a probe
photon colliding with two plane waves (a low-frequency wave and a
high-frequency wave), even for the case of the field strength
parameter $\xi_{1,2} < 1$.

In addition, we want to point out here that as shown in
Figs.~\ref{spectra40} and \ref{spectrum40s}, the spectra $d
W_{n_1n_2}/d q_p$ become large values around the maximal values
$q_{p}^{max}(n_1,n_2)$ for fixed values of $(n_1,n_2)$. This is
mainly due to the factor $1/\sqrt{u(u-1)}$ in the pair-production
probability of Eqs.~(\ref{ppptv}), (\ref{pppov}) and
(\ref{dppptwn}). The integrations of Eqs.~(\ref{ppptv}) and
(\ref{pppov}) over spectra $d W/d q_p$ are finite, i.e., the total
pair-production probability $W$ is finite.

\section{Summary and remarks}
\label{sec:sum}

Based on the Volkov solutions of the Dirac equation in two plane
waves, we studied the nonlinear Breit-Wheeler process of pair
production off a probe photon colliding with a low-frequency and a
high-frequency plane wave that propagate in the same direction. We
analyzed the difference between the nonlinear Breit-Wheeler process
(\ref{two-wave}) of $e^-~e^+$ pair production off a probe photon
colliding with two plane waves and the process (\ref{one-wave}) of
pair production off the probe photon colliding with each of these
two plane waves.

The results show that the high multiphoton phenomenon is clearly
evident in the nonlinear Breit-Wheeler process of pair production
off a probe photon colliding with a low-frequency and a
high-frequency plane wave. We also show the phenomenon of
Breit-Wheeler pair production with simultaneous photon emission into
the low-frequency wave by absorbing one photon from the
high-frequency wave and the probe photon. This phenomenon of pair
production with simultaneous photon emission into the wave cannot
happen in the nonlinear Breit-Wheeler process of pair production off
a probe photon colliding with one plane wave only. In the case of
the electromagnetic plane waves of intensities $I_1 = I_2 =
10^{18}~\rm{W/cm^2}$, the frequency $\omega_1\sim {\rm{eV}}$, and
$\omega_2 = 100~\omega_1$, i.e., $\xi_1<1$ and $\xi_2\ll 1$, the
contributions to the pair-production probability from the process
with one photon ($n_2=1$) absorbed from the high-frequency wave and
a large number ($|n_1|$) of photons absorbed from (or emitted into)
the low-frequency wave are still significantly large even at
$|n_1|\simeq 20$. As a comparison, in the absence of the
high-frequency wave, the contributions to the pair-production
probability from the processes can be negligible when the number of
photons absorbed from the low-frequency wave is larger than $5$,
i.e., $n_1> 5$. This indicates that the multiphoton phenomenon
cannot be evident in the presence of one plane wave only with field
strength parameter $\xi_{1,2} <1$. This means that the presence of
the high-frequency wave enhances the contributions of the
low-frequency wave to the pair-production probability via the high
multiphoton processes. We also present the spectra (multipeak
structure) of the created $e^-~e^+$ pair to show the effects of the
phenomenon of pair production with simultaneous photon emission into
the low-frequency wave and high multiphoton (absorption and
emission) phenomenon in this two-wave process. These phenomena can
be studied by using the already available technology of lasers even
for the wave field strength parameter $\xi_1 <1$.

To end this article, we would like to mention that it would be
interesting to make a systematic analysis of the nonlinear
Breit-Wheeler processes of $e^-~e^+$ pair production off a probe
photon colliding with two plane wave fields in the large range of
parameters $\xi_1$, $\xi_2$, $\zeta_1$, and $\zeta_2$ of two plane
waves and high-energy photons.

\vspace{0.5cm}
\noindent %
{\bf Acknowledgements}

Authors are grateful to Prof.~ Ruffini for his support and
encouragement to work on the physics of strong fields. We are also
grateful to the anonymous referee for his/her suggestions and
comments, which helped us to improve our article. Yuan-Bin Wu is
supported by the Erasmus Mundus Joint Doctorate Program through
Grant Number 2011-1640 from the EACEA of the European Commission.



\end{document}